\documentclass[aps,pra,preprint,showpacs,superscriptaddress]{revtex4} 
\usepackage{amssymb}
\usepackage[dvips]{epsfig}
\usepackage{subfigure}
\usepackage{amsmath}

\newcommand{\scl}{0.2} 

\begin{document}

\title{Stability Through Asymmetry: \\
Modulationally Stable Nonlinear Supermodes of Asymmetric non-Hermitian Optical Couplers}

\author{Yannis Kominis}
\affiliation{School of Applied Mathematical and Physical Science, National Technical University of Athens, Athens, Greece}
\author{Tassos Bountis}
\affiliation{Department of Mathematics, School of Science and Technology, Nazarbayev University, Astana, Republic of Kazakhstan}
\author{Sergej Flach}
\affiliation{Center for Theoretical Physics of Complex Systems, Institute for Basic Science, Daejeon, Korea}

\begin{abstract}
We analyze the stability of a non-Hermitian coupler with respect to
modulational inhomogeneous perturbations in the presence of unbalanced gain and loss.
At the parity-time (PT) symmetry point the coupler is unstable. Suitable symmetry breakings lead to
an asymmetric coupler, which hosts nonlinear supermodes. A subset of these broken symmetry cases finally
yields nonlinear supermodes which are stable against modulational perturbations. The lack of symmetry requirements
is expected to facilitate experimental implementations and relevant photonics applications.
\end{abstract}

\maketitle

\section{Introduction}

Modulational Instability (MI) is one of the most ubiquitous types of instability in natural and technological systems and is considered as one of the most fundamental nonlinear processes in many different branches of modern physics, including fluid dynamics, electrodynamics, plasma physics and nonlinear optics. \cite{ZO_09} MI is a symmetry-breaking instability under which, a constant-intensity plane wave (CW) is spontaneously modulated to a spatially or temporally inhomogeneous pattern due to the interplay between the diffraction or dispersion and self-phase modulation (SPM) nonlinearity. In the spatial domain this effect results in a wave breaking up into filaments. In the temporal domain it transforms a CW into a pulse train. Early studies of the MI date back to the 1960s in the context of fluid dynamics and water waves \cite{W_65, BF_67} as well as in the context of nonlinear  electromagnetic wave propagation 
\cite{O_67}.

The universality of the MI resulted in extended studies under more general settings. For coupled nonlinear waves as in the case of Cross-Phase Modulation (XPM), occurring for example in birefrigent nonlinear optical fibers, the concept has been generalized to the Vector Modulation Instability (VMI) under which the constant polarization components of the wave can be simultaneously destabilized due to the presence of random noise perturbations \cite{BZ_70, Agrawal}.  
Another important generalization takes place in active systems with energy pumping and loss. Under proper balance conditions a CW can be formed as an equilibrium that can be either modulationally stable or unstable. Such non-conservative systems have been studied in terms of the Ginzburg-Landau (GL) equation \cite{ECKHAUS_65, NEWELL_69, NEWELL_74} for the case of homogeneous gain and loss mechanisms.

On the other hand, nonlinear wave formation and propagation in media with spatially inhomogeneous gain and loss has been a subject of recent intense research interest in the field of non-Hermitian photonics. Originally motivated by quantum physics studies of complex potentials with a real spectrum under parity-time ($\mathcal{PT}$) symmetric conditions \cite{B_07,M_11}, the field of non-Hermitian photonics has rapidly grown due to a large number of wave phenomena that have no counterpart in conventional Hermitian photonic structures \cite{PT_1, PT_2, PT_3, PT_4, PT_5, PT_6}. Theoretical and experimental studies have been focused almost exclusively to the case of $\mathcal{PT}$-symmetric structures, where gain and loss are exactly balanced, and interesting applications have been presented, including perfect optical absorption \cite{absorber_1, absorber_2}, non-reciprocal light transmission \cite{Feng_11, Bender_13}, unidirectional invisibility \cite{Lin_11, Feng_13}, nonlinear switching \cite{Sukhorukov_10}, as well as microcavity lasing \cite{Peng_14, lasers_1, lasers_2}.

Wave formation and propagation, in inhomogeneous media, is governed either by coupled mode equations (under a tight-binding approximation) or by partial differential equations (PDEs) with spatially-varying coefficients, which, in general, do not admit constant-intensity wave solutions. Therefore, the spatial inhomogeneity prevents the existence of spatially uniform constant-intensity waves, so that any study on MI is meaningless. One way to overcome this difficulty and preserve the existence of constant-intensity waves, consists in introducing specific types of inhomogeneities. Recent attempts, along this direction, have utilized a rather special complex potential, in which the spatial forms of the real and the imaginary part are functionally related.  Under this condition, constant-intensity waves with non-trivial spatially inhomogeneous phase are shown to exist \cite{Makris_15, Cole_16}. However, this condition restricts the generality of the structure and its potential application in realistic configurations. Moreover, it is shown that the respective constant-intensity waves are modulationally unstable either for self-focusing or for self-defocusing nonlinearities. 

 In this work, we overcome the difficulty of the existence of a constant-intensity wave in an inhomogeneous non-Hermitian structure by considering a system of coupled waveguides, one with gain and one with loss. As recently shown, such a structure has constant-intensity Nonlinear Supermodes (NS) consisting of different wave amplitudes in each waveguide. \cite{SREP} The spatially asymmetric NS are shown to exist only when the system is not $\mathcal{PT}$-symmetric, that is when gain and loss are not exactly balanced and the two waveguides (or their respective guided modes) are not identical. Therefore, the formation of the NS does not result from a direct gain-loss balance, but from a more complex dynamical balance between all the mechanisms governing wave formation that is, linear coupling, nonlinearity, waveguide asymmetry and gain/loss inhomogeneity.  It is worth noting that, in the $\mathcal{PT}$-symmetric case, no asymmetric finite-power modes exist, so that directed power transport capabilities are quite restricted \cite{Ramezani_10}. 
 
 The Vector Modulation Instability of the NS of the asymmetric coupler is studied in terms of the coupled mode equations of the system, and the roles of gain-loss inhomogeneity and waveguide asymmetry are investigated. More importantly, it is shown that there exist extended parameter regions where these asymmetric Nonlinear Supermodes are modulationally stable. 

 \begin{figure}[!h]
  \begin{center}
  \subfigure[]{\scalebox{\scl}{\includegraphics{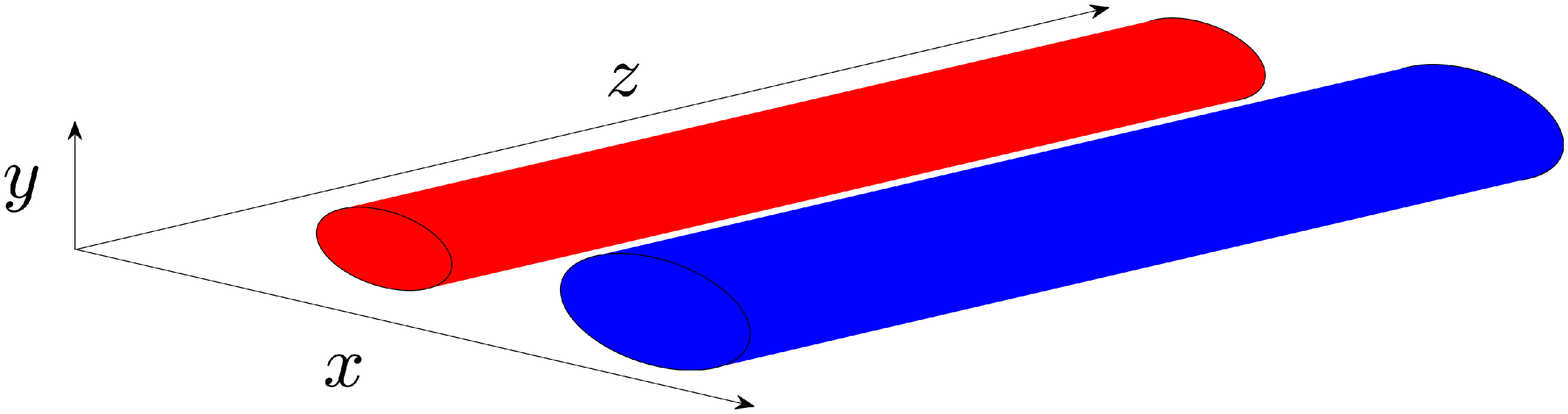}}}
  \subfigure[]{\scalebox{\scl}{\includegraphics{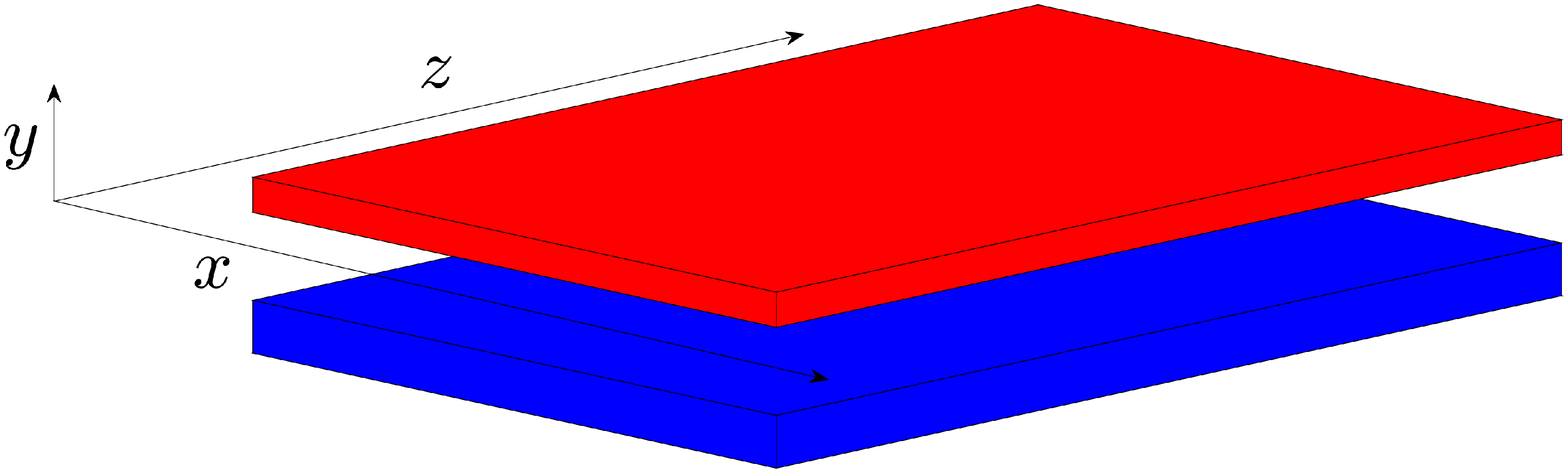}}}
  \caption{(a) Fiber coupler. The wave profiles in the spatial dimensions $x$ and $y$ are determined by the waveguided modes of the two fibers. The ``transverse'' variable $t$ [Eq. (\ref{CM})] refers to the time dependence of the wave. (b) Slab waveguide coupler. The wave profiles in the spatial dimension $y$ are determined by the waveguided modes of the two slabs and their time dependence is harmonic (steady state). The ``transverse'' variable $t$ [Eq. (\ref{CM})] refers to the wave profile along the infinite spatial $x$ dimension ($t \equiv x$).  }
  \end{center}
\end{figure}

\section{Coupled Mode Equations}
The model governing the evolution of the electric field envelopes in the two waveguides consists of two coupled mode equations having the form of a Vector NonLinear Schr\"odinger (VNLS) system with gain and loss:  
\begin{eqnarray}
 i\frac{\partial u_{1,2}}{\partial z}&+&\frac{\partial^2 u_{1,2}}{\partial t^2}+(\beta_{1,2}+i\alpha_{1,2})u_{1,2}  \nonumber \\
 &+&\gamma\left(|u_{1,2}|^2+\sigma|u_{2,1}|^2\right)u_{1,2}+\frac{\kappa}{2}u_{2,1}=0 \label{CM} 
\end{eqnarray}
where $\beta_j+i\alpha_j$ is the complex propagation constant of waveguide $j$ with $a_j>0$  ($a_j<0$) referring to loss (gain), $\kappa/2$ is the linear coupling coefficient and $\gamma$, $\sigma$ are the nonlinear SPM and XPM parameters, respectively \cite{Chen_92}. The propagation distance is denoted by $z$ and the ``transverse'' variable $t$ denotes real time for the case of pulse propagation in coupled fibers, or a transverse spatial dimension ($t \equiv x$) for the case of beam propagation in coupled slab waveguides, depicted in Figs. 1(a) and (b) respectively. As shown in Ref. \cite{SREP}, the system supports two Nonlinear Supermodes (NS) of the form $u_i=A_i\exp(ibz)$, $(i=1,2)$ with constant complex amplitudes $A_i=|A_i|\exp(i\phi_i)$  ($\partial A_i/\partial t = \partial A_i/\partial z = 0$). These amplitudes of the two NS are given by
\begin{eqnarray}
 |A_1|^2 &=& \frac{\beta_1 \alpha (\beta-1)+\frac{\kappa}{2}\sqrt{\alpha}(\alpha-1)\cos\phi}{\gamma(1-\sigma)(\alpha-1)} 
\; , \\
 |A_2|^2 &=& \frac{1}{\alpha}|A_1|^2 \; .
\end{eqnarray}
The propagation constant $b$ and the phase difference $\phi\equiv \phi_1-\phi_2$ are defined as
\begin{equation}
 b=\frac{1}{2}\left[\beta_1(\beta+1)+\frac{\kappa}{2}\frac{\alpha+1}{\sqrt{\alpha}}\cos\phi
 +\gamma(\sigma+1)\frac{\alpha+1}{\alpha}|A_1|^2\right] \;,
\end{equation}
\begin{equation}
 \sin\phi=\frac{2|\alpha_1|\sqrt{\alpha}}{\kappa} \; ,
\end{equation}
where $\alpha \equiv -\alpha_2/\alpha_1$ and $\beta \equiv \beta_2/\beta_1$.
The parameter conditions for the existence of the NS are 
\begin{equation}
 0<\alpha<\left(\frac{\kappa}{2\alpha_1}\right)^2 \;, \label{condition_a}
\end{equation}
\begin{equation} 
\gamma(1-\sigma)(\alpha-1)\alpha\beta_1 \left[ \beta-1 \pm \frac{|\alpha_1|}{\beta_1}\frac{\alpha-1}{\sqrt{\alpha}} \sqrt{\left(\frac{\kappa}{2\alpha_1}\right)^2-\alpha}\right]>0 \; , \label{condition_b}
\end{equation}
with the different signs corresponding to the two Nonlinear Supermodes. The first condition ($\alpha>0$) necessitates the presence of both gain and loss in the structure.  The second one determines the number of NS so that, depending on the parameters of the system, there exist either zero, one or two NS with equal amplitude ratios and unequal phase differences. The existence of Nonlinear Supermodes requires that the couplers are nonlinear ($\gamma\neq 0$), and
with asymmetry in the SPM and XPM coefficients ($\sigma \neq 1$) and that the system is not $\mathcal{PT}-$symmetric ($\alpha \neq 1$). \cite{SREP}
In order to study the Modulation Instability (MI) of the system we consider perturbations of the form
\begin{equation}
 u_{1,2}=\left(A_{1,2}+\epsilon_{1,2}\right)e^{ibz} \;.  \nonumber
\end{equation}
We obtain the following linear equation system
\begin{eqnarray}
 i\frac{\partial \epsilon_{1,2}}{\partial z}+\frac{\partial^2 \epsilon_{1,2}}{\partial t^2}+F_{1,2}\epsilon_{1,2}+H_{1,2}\epsilon_{1,2}^*+M_{1,2}\epsilon_{2,1}+G\epsilon_{2,1}^* &=&0 \; ,
\end{eqnarray}
with
\begin{eqnarray}
 F_{1,2} &=& \beta_{1,2}-b+i\alpha_{1,2}+\gamma\left(2|A_{1,2}|^2+\sigma|A_{2,1}|^2\right) \;,\\
 H_{1,2} &=& \gamma A_{1,2}^2 \; , \\
 M_1&=&M_2^*=M= \gamma A_1 A_2^*+\frac{\kappa}{2} \; , \\
 G &=& \gamma\sigma A_1 A_2 \; .
\end{eqnarray}
Without loss of generality we consider perturbations consisting of Fourier modes
\begin{equation}
 \epsilon_{1,2}=c_{1,2}e^{i(\omega t-\lambda z)}+d_{1,2}^*e^{-i(\omega t-\lambda^*z)} \; , 
\end{equation}
where $\omega$ is the frequency and $\lambda$ the (complex) wavenumber of the perturbation. We then obtain the linear eigenvalue problem 
\begin{equation}
 L\psi=\lambda\psi
\end{equation}
with the eigenvector $\bold{\psi}=[c_1,d_1,c_2,d_2]$ and the matrix
\begin{equation}
L=\left[
  \begin{array}{cccc}
   \omega^2-F_1 	&	-H_1      	&	-M 	&	-G  		\\
      H_1^*     	&-\omega^2+F_1^*	&	 G^*	&	M^*		\\
      -M^*		&	-G		&\omega^2-F_2	&	-H_2		\\
      G^*		&	M		&	H_2^*	& -\omega^2+F_2^* 
  \end{array}
  \right] \;.
\end{equation}
Modulational instability takes place for perturbations with an eigenvalue $\lambda$ with a positive imaginary part, with the latter corresponding to the growth rate of the unstable mode. We are searching for stable propagation cases when the
eigenvalues have negative imaginary parts only.

\section{Results and Discussion}
It is well known that for the standard conservative scalar NonLinear Schr\"odinger (NLS) equation, MI occurs for self-focusing nonlinearity (or anomalous dispersion regime in fibers)  whereas CW are modulationally stable for self-defocusing nonlinearities (or normal dispersion regime in fibers). The sign of the nonlinearity crucially determines the interplay between the linear diffraction and the nonlinearly induced phase shifts resulting either in stability or instability \cite{Agrawal}.
In the following, we investigate the role of gain and loss inhomogeneity as well as of the asymmetry of the system on the MI of the Nonlinear Supermodes and study its dependence on relevant parameters of the configuration. Without loss of generality, we set the linear coupling coefficient $\kappa=1$ as well as the propagation constant and gain/loss coefficients of the first waveguide $\beta_1=1$, $\alpha_1=1$ and investigate MI in terms of the parameters related to the asymmetry of the system ($\beta$, $\alpha$) and the nonlinearity ($\gamma$,$\sigma$). The instability is studied in terms of the growth rate ($g$) corresponding to the maximum imaginary part of the eigenvalues
\begin{equation}
g=\max_{i=1,2,3,4}\left(\mbox{Im}\{\lambda_i\}\right) 
\end{equation}
with $g>0$ being the condition for MI.

\begin{figure}[!h]
  \begin{center}
  \subfigure[]{\scalebox{\scl}{\includegraphics{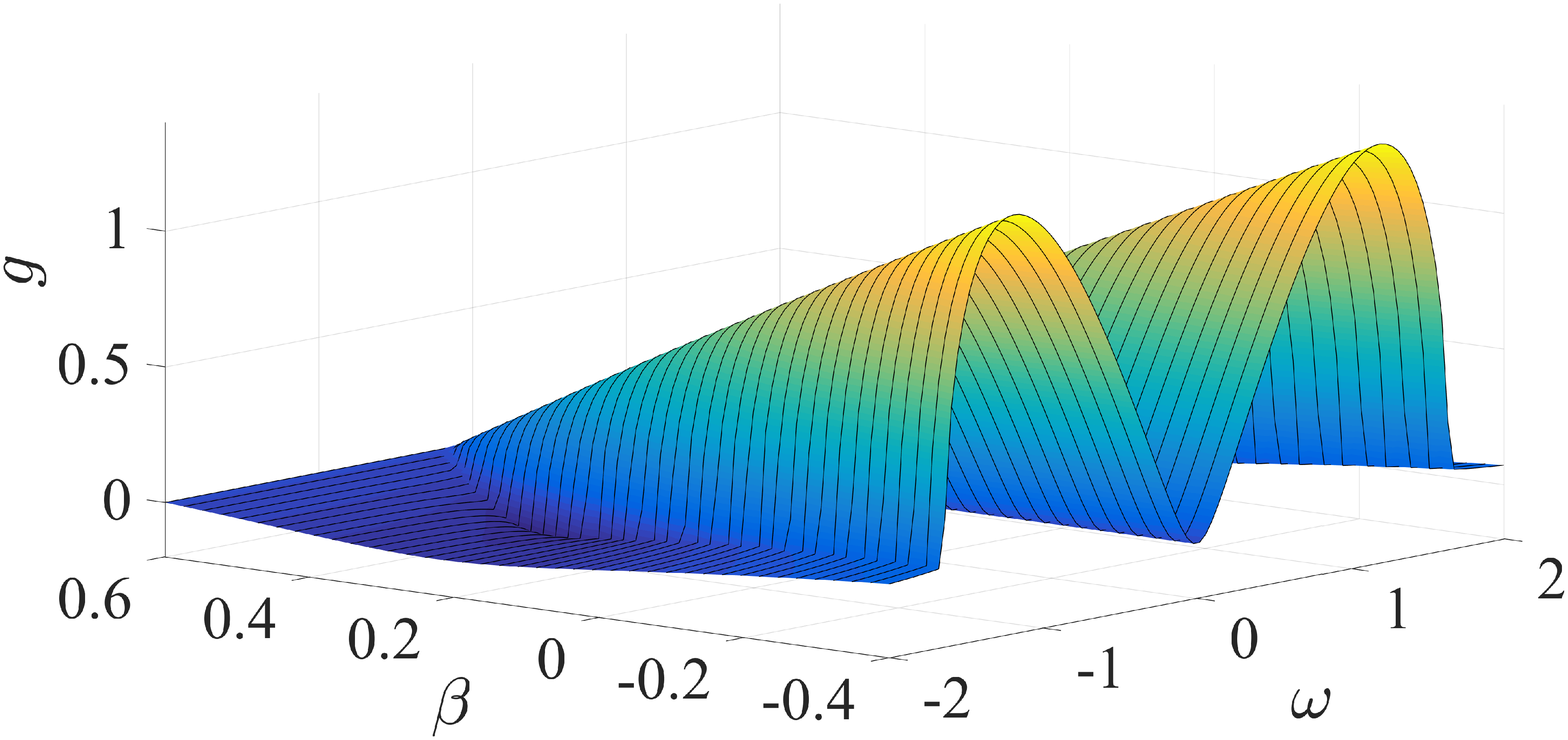}}}
  \subfigure[]{\scalebox{\scl}{\includegraphics{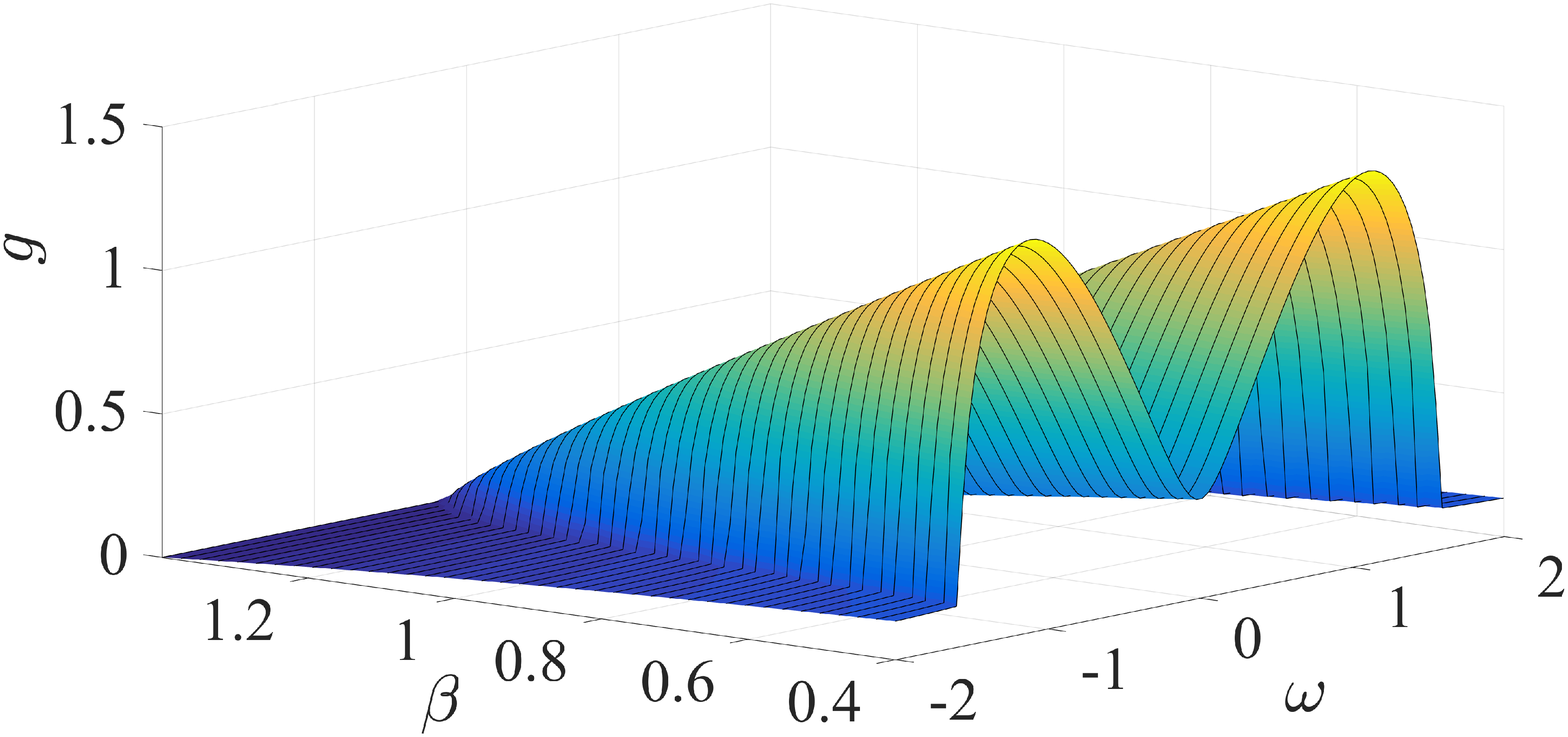}}}
  \caption{Growth rate $g$ for the case of self-focusing nonlinearity ($\gamma=1$) and no XPM ($\sigma=0$) for a gain-loss imbalance corresponding to $\alpha=0.2$ for the two Nonlinear Supermodes (a) and (b). Both NS are modulationally unstable.}
  \end{center}
\end{figure}

\begin{figure}[!h]
  \begin{center}
  \subfigure[]{\scalebox{\scl}{\includegraphics{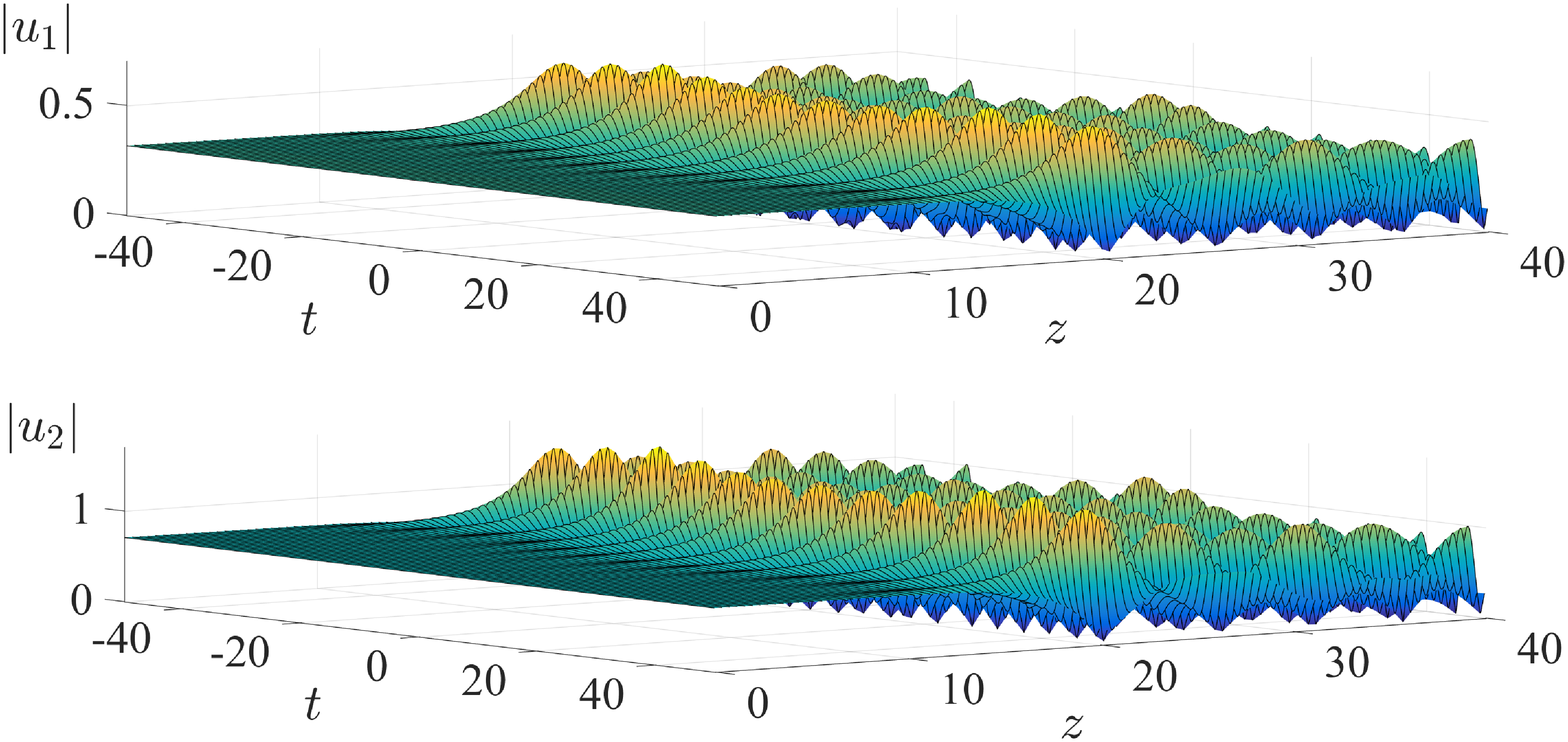}}}
  \subfigure[]{\scalebox{\scl}{\includegraphics{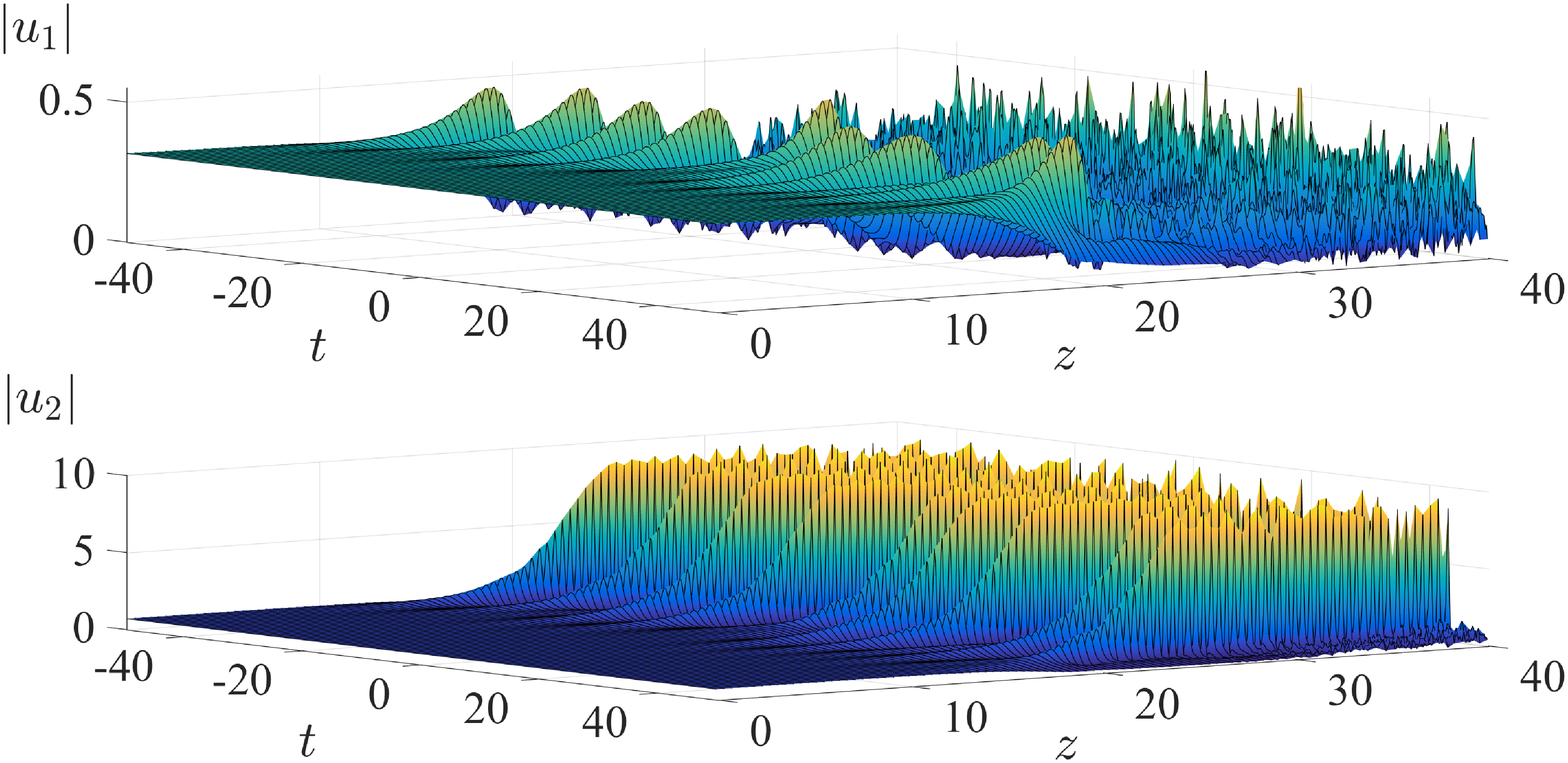}}}
  \caption{Development of the Modulation Instability of the Nonlinear Supermodes under a random noise perturbation of the order of $10^{-2}$ superimposed at $z=0$, for the cases corresponding to Fig. 2(a) with $\beta=0.2$ and Fig. 2(b) with $\beta=1$, respectively.}
  \end{center}
\end{figure}

For the case of self-focusing nonlinearity ($\gamma=1$) and no XPM ($\sigma=0$), we always find positive values of $g$ for both Nonlinear Supermodes, as shown in Fig. 2 for a gain and loss imbalance corresponding to $\alpha=0.2$ and different values of $\beta$. One of the Nonlinear Supermodes [Fig. 2(a)] is stable under homogeneous perturbations ($\omega$=0), whereas the other [Fig. 2(b)] is not, in agreement with previous stability analysis of this restricted case  \cite{SREP}. The range of $\beta$ is the semi-infinite interval determined from the existence conditions of the Nonlinear Supermodes (\ref{condition_b}) and the growth rate of the instability ($g$) uniformly increases with decreasing $\beta$. The dynamical evolution of MI is studied by adding small random noise perturbations of the order of $10^{-2}$ to the exact Nonlinear Supermode solution at $z=0$, as shown in Figs. 3(a) and (b) for both Nonlinear Supermodes with $\beta=0.2$ and $\beta=1$, respectively. For the self-focusing case ($\gamma=1$) a nonzero XPM coefficient ($\sigma\neq0$)   can only alter the magnitude and not the sign of the growth rate and does not result in qualitative changes of the modulation instability of the Nonlinear Supermodes. This is similar to the case of Vector Modulational Instability in conservative VNLS systems with no gain and loss where the NS are always modulationally unstable for the self-focusing case irrespectively of the XPM coefficient \cite{Agrawal}.

\begin{figure}[!h]
  \begin{center}
  \subfigure[]{\scalebox{\scl}{\includegraphics{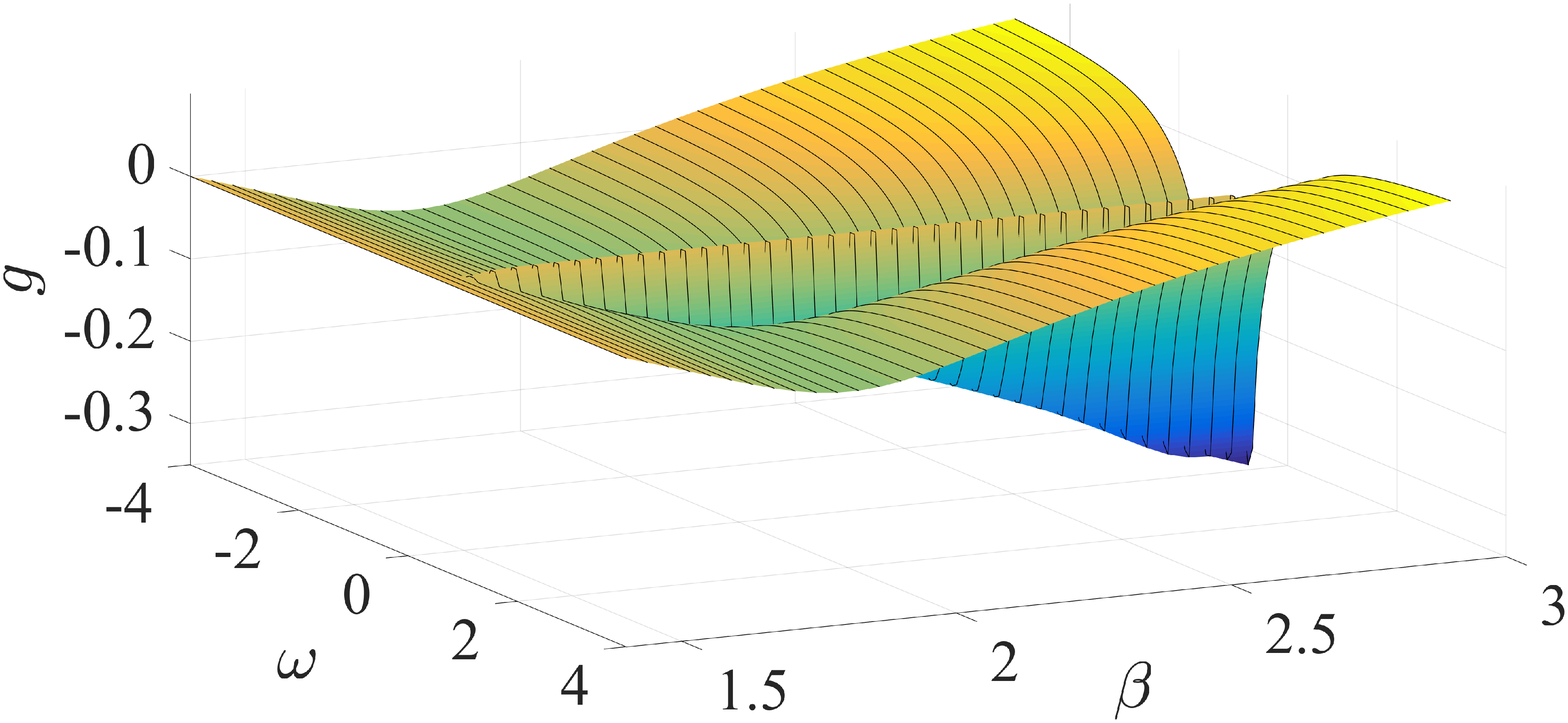}}}
  \subfigure[]{\scalebox{\scl}{\includegraphics{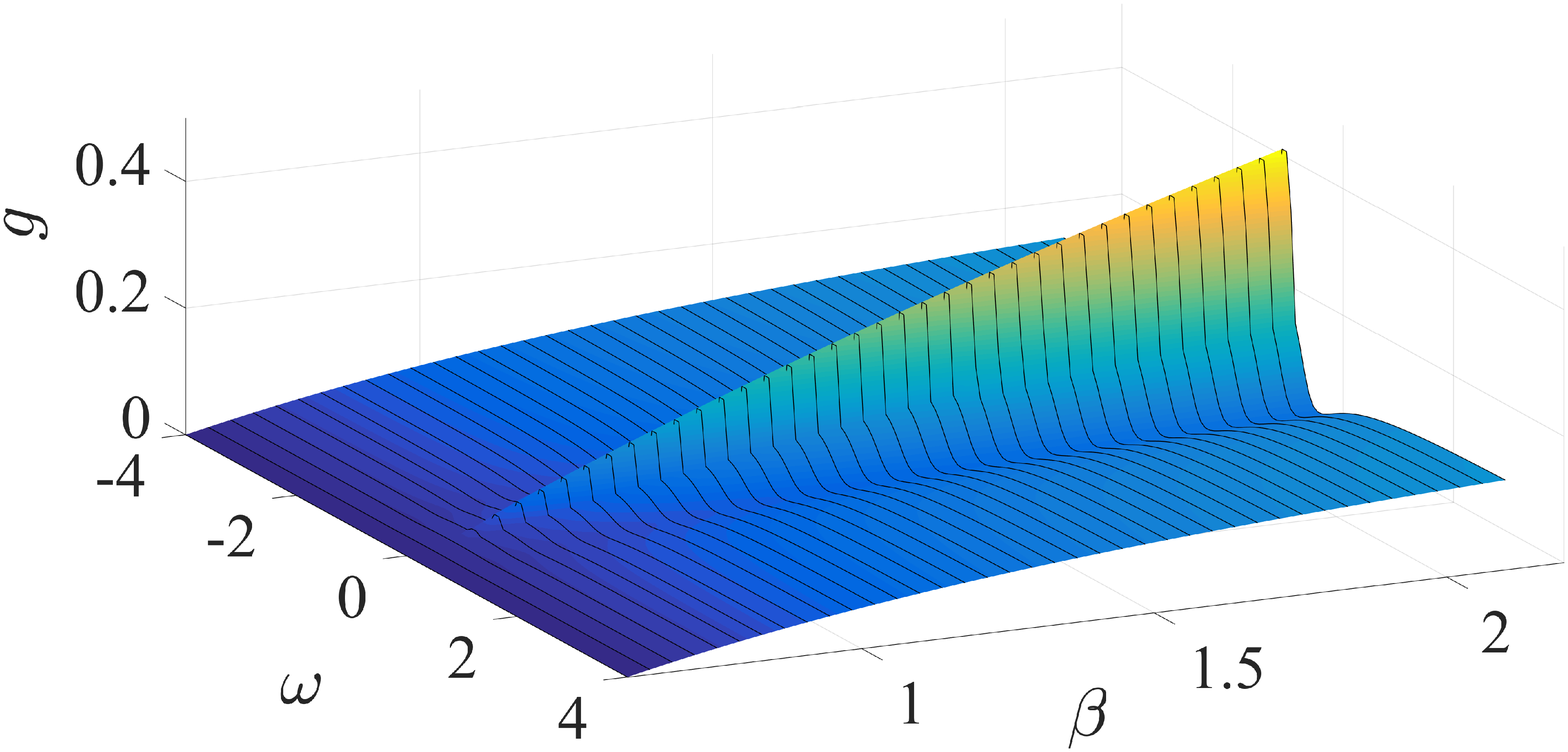}}}
  \caption{Growth rate $g$ for the case of self-defocusing nonlinearity ($\gamma=-1$) and no XPM ($\sigma=0$) for a gain-loss imbalance corresponding to $\alpha=0.2$ for the two Nonlinear Supermodes (a) and (b). The first NS (a) is shown to be modulationally stable ($g<0$ for all $\omega$) for a range of parameter $\beta$ values. The second NS (b) is always modulationally unstable.}
  \end{center}
\end{figure}

\begin{figure}[!h]
  \begin{center}
  \subfigure[]{\scalebox{\scl}{\includegraphics{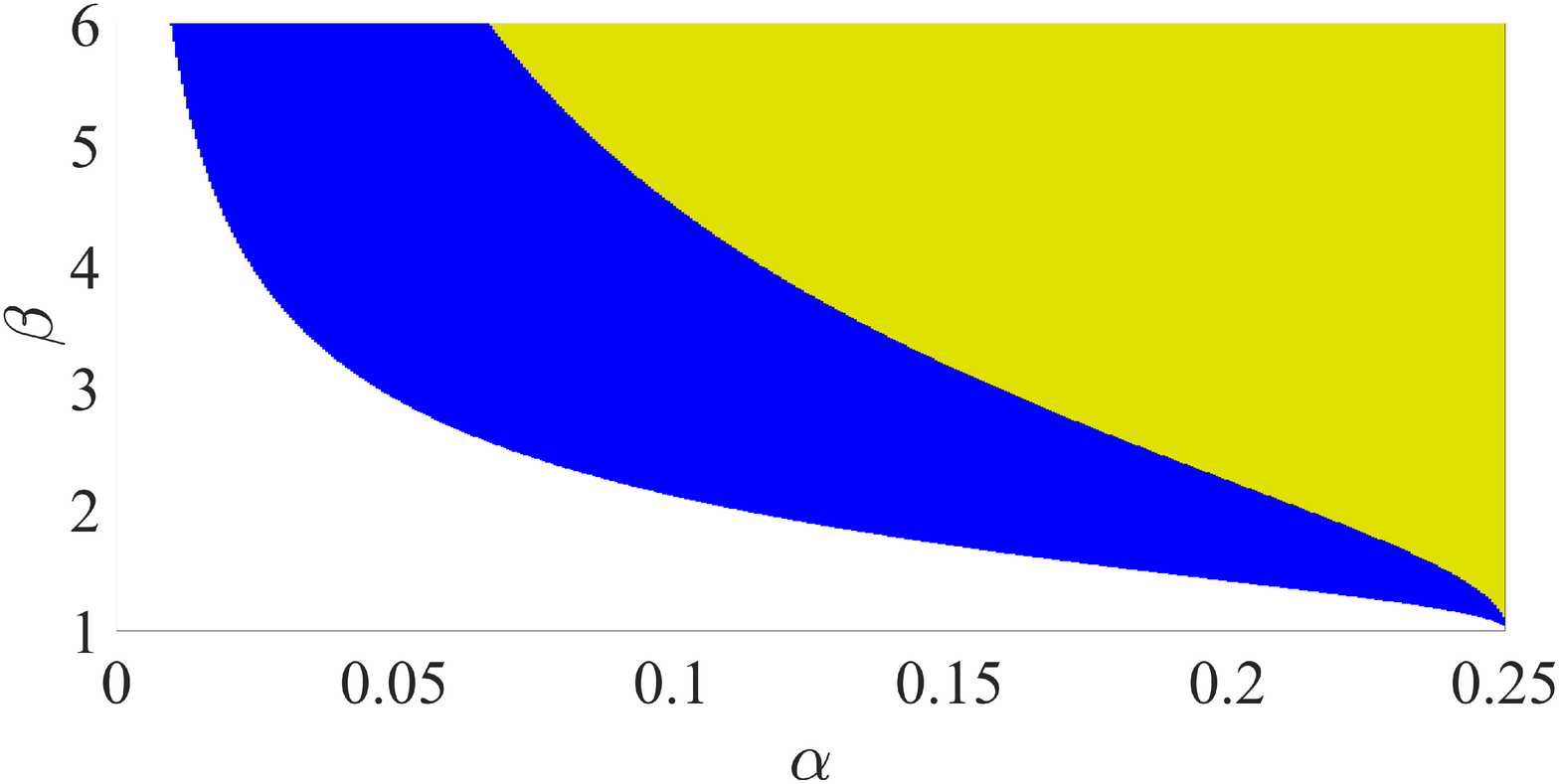}}}
  \subfigure[]{\scalebox{\scl}{\includegraphics{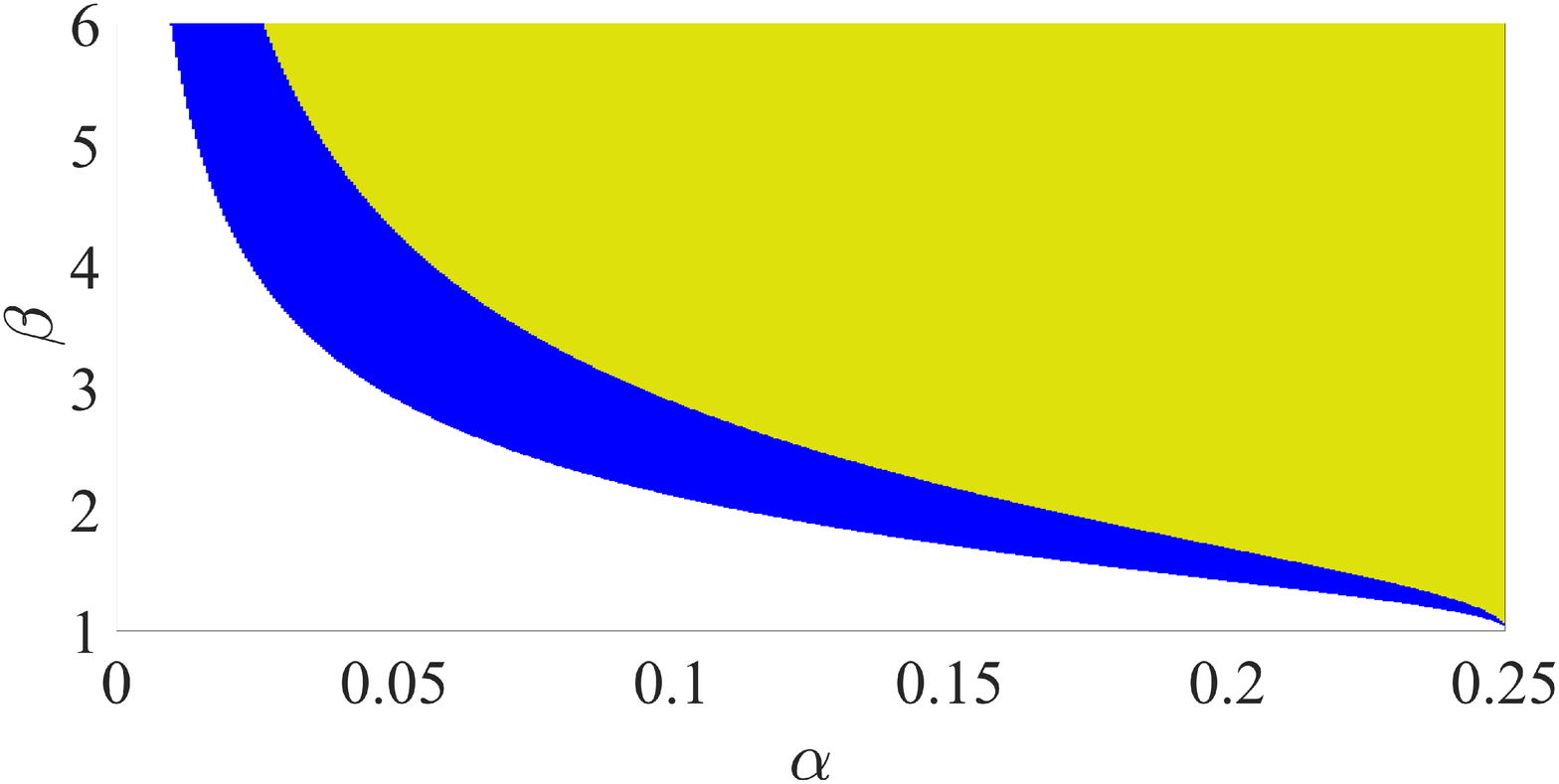}}}
  \caption{Existence domain of modulationally stable Nonlinear Supermodes in terms of the parameters $\alpha$ and $\beta$ expressing the asymmetry of the structure. The Nonlinear Supermodes are stable against homogeneous perturbations (with $\omega=0$) in the whole shaded area and modulationally stable (against perturbations with arbitrary $\omega$) in the blue-shaded area. Cases of zero ($\sigma=0$) and nonzero ($\sigma=0.5$) XPM coefficients are shown in (a) and (b), respectively.} 
  \end{center}
\end{figure}

In contrast to the standard scalar NLS equation, in conservative VNLS systems, constant-intensity CWs can be destabilized due to coupling even for self-defocusing nonlinearities \cite{Agrawal}. In the following, we investigate the parameter space of the asymmetric non-Hermitian coupler for the existence of modulationally stable Nonlinear Supermodes for a self-defocusing nonlinearity ($\gamma=-1$).  First we consider a case with no XPM ($\sigma=0$) and gain and loss imbalance corresponding to $\alpha=0.2$. The growth rate $g$ for each perturbation frequency $\omega$ and different values of $\beta$ for the two NS is depicted in Fig. 4. We observe that one of the Nonlinear Supermodes [Fig. 4(a)] is modulationally stable ($g<0$ for all $\omega$) for a part of its domain of existence with respect to parameter $\beta$, whereas the second Supermode [Fig. 4(b)] is always  modulationally unstable. The existence domain of modulationally stable Nonlinear Supermodes in the control parameter space of the asymmetry parameters $\alpha$ and $\beta$ is shown in Fig. 5(a). The Nonlinear Supermodes are stable under homogenous perturbations $\omega=0$ in the whole shaded area of the parameter space; and they are modulationally stable, under perturbations with arbitrary $\omega$, in the blue-shaded regions. In Fig. 5(b), it is shown that a non-zero XPM coefficient ($\sigma=0.5$) results in a narrower domain of modulation stability in comparison to a zero one, shown in Fig. 5(a). This is similar to the case of a conservative VNLS system where the XPM is shown to have a destabilizing effect \cite{Agrawal}. In both cases of different $\sigma$, it is clear that a smaller $\alpha$, corresponding to larger gain and loss imbalance between the two waveguides, results in a wider domain of modulational stability. Therefore, there is enough freedom in parameter selection for the control of the existence conditions for a modulationally stable Nonlinear Supermode. 

\begin{figure}[!h]
  \begin{center}
  \subfigure[]{\scalebox{\scl}{\includegraphics{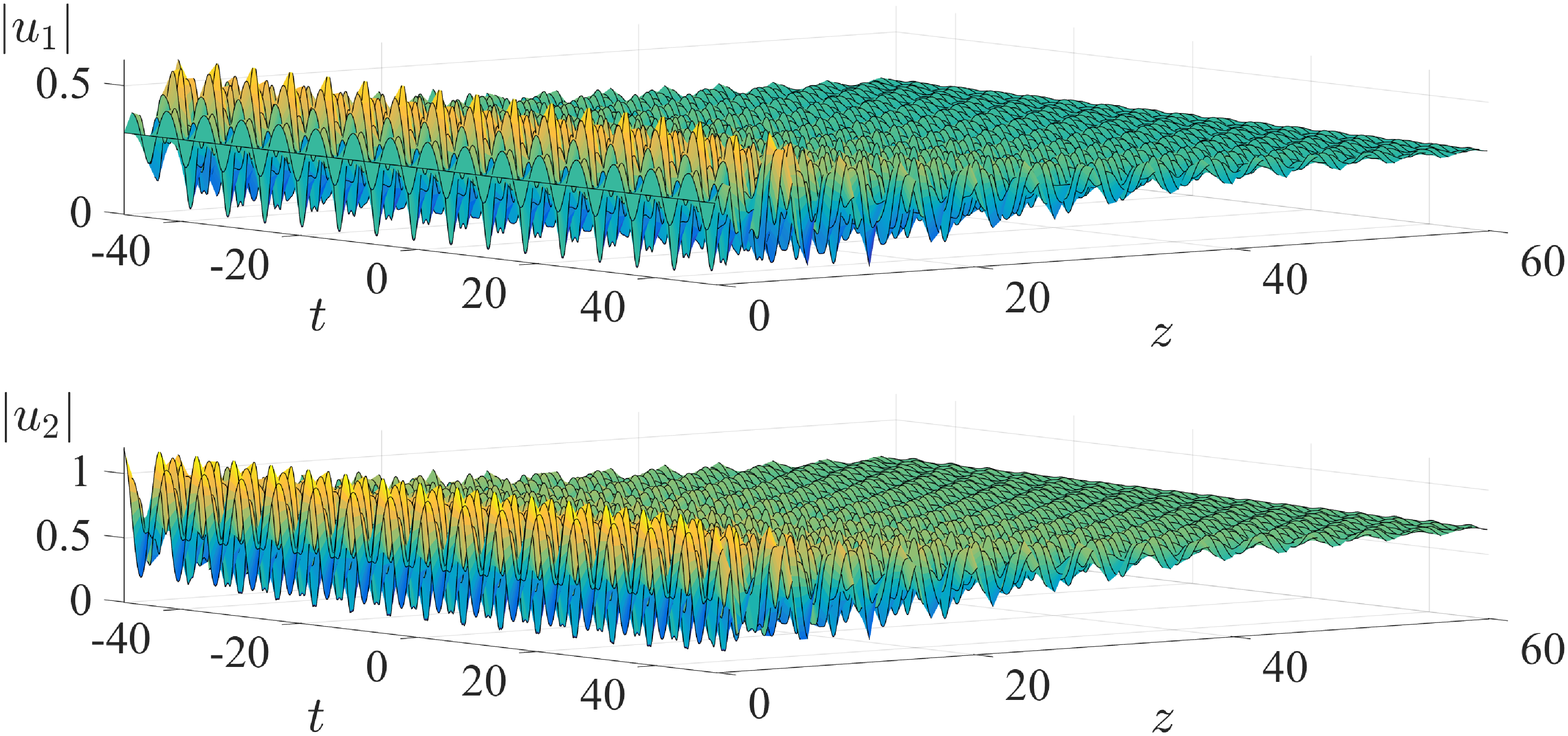}}}
  \subfigure[]{\scalebox{\scl}{\includegraphics{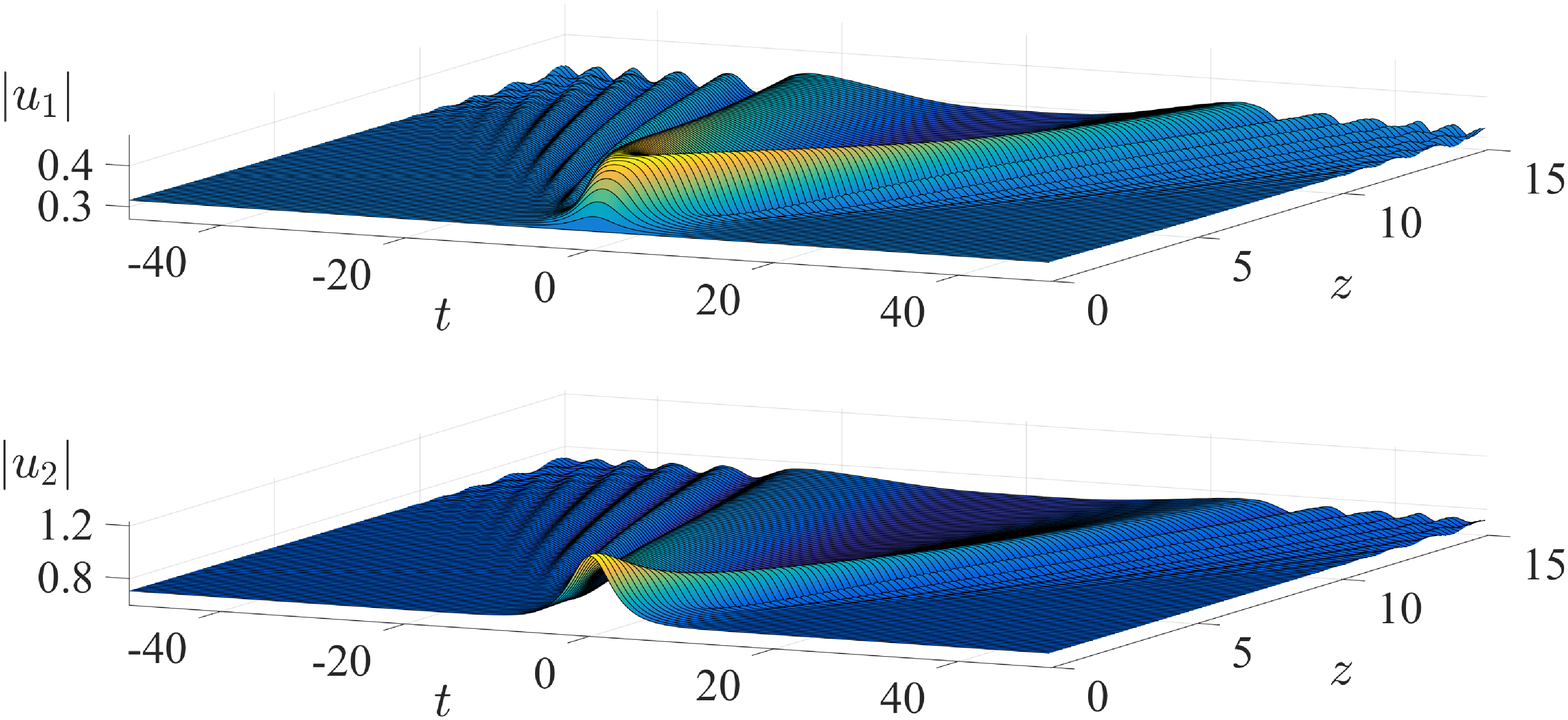}}}
  \caption{Wave evolution of periodic and localized perturbations superimposed on a modulationally stable Nonlinear Supermode. Parameter values as in Fig. 4(a), and $\beta=1.8$. In the second waveguide, a periodic wave $\tilde{u}=0.5 \cos(t)$ (a) and a localized wave $\tilde{u}=0.5\mbox{sech}(0.5t)$ (b) have been superimposed on the stable NS at $z=0$. }
  \end{center}
\end{figure}

The modulational stability of the respective NS results in the asymptotic decay (due to negative growth rates) of any type of finite perturbation. The wave evolution for a spatially (temporally) periodic and a localized excitation superimposed on such a modulationally stable Nonlinear Supermode at $z=0$ is depicted in Figs. 6(a) and (b), respectively. The parameters of the system correspond to the case shown in Fig. 4(a) with $\beta=1.8$. In both cases, it is clear that the initial excitations evolve to the stable NS. Therefore, dissipative dynamics takes place on a constant background that is not itself excited either by periodic or by localized waves. Periodic perturbations are uniformly dissipated, whereas localized perturbations, having the form of a beam (pulse), split in secondary beams (pulses) that are subsequently dissipated. In both cases, the system evolves to the asymptotically stable NS. 

\section{Conclusions}
The Modulation Instability of an inhomogeneous non-Hermitian photonic structure is studied for the fundamental paradigm of two coupled waveguides. This structure is shown to support nontrivial Nonlinear Supermodes due to its asymmetric characteristics. The existence and the modulation instability of the NS are investigated in the parameter space of the system. It was shown that modulationally stable Nonlinear Supermodes can exist for a wide parameter range, as a result of the asymmetry of the system. The generic configuration is not restricted by any symmetry conditions and facilitates experimental implementation and interesting photonics applications utilizing the existence of stable continuous waves for directed power transport and/or their controllable breaking either to time periodic pulse trains or to spatial beam filaments.

\section*{Acknowledgements}

Y. K. is grateful to the School of Science and Technology (SST) of Nazarbayev University, Astana, Kazakhstan for its hospitality during his visit at NU in November, 2016, and thanks Prof. V. Kovanis of the Physics Department of SST for many useful conversations on topics related to the content of the present paper. T.B. gratefully acknowledges partial support of Nazarbayev University through his ORAU grant 2017 - 2020. This work was supported by the Institute for Basic Science through Project Code IBS-R024-D1.

%
%


\end{document}